\documentclass{PoS}\usepackage{amsmath}\title{Local-Duality QCD Sum
Rules for Pion Elastic and $(\pi^0,\eta,\eta')\to\gamma\,\gamma^*$
Transition Form Factors Revisited}\ShortTitle{Pion Elastic and
$(\pi^0,\eta,\eta')\to\gamma\,\gamma^*$ Transition Form Factors}
\author{\speaker{Irina Balakireva}\\D.~V.~Skobeltsyn Institute of
Nuclear Physics, Moscow State University, 119991, Moscow, Russia\\
E-mail: \email{iraxff@mail.ru}}\author{Wolfgang Lucha\\Institute
for High Energy Physics, Austrian Academy of Sciences,
Nikolsdorfergasse 18, A-1050 Vienna, Austria\\E-mail:
\email{Wolfgang.Lucha@oeaw.ac.at}}\author{Dmitri Melikhov\\
Institute for High Energy Physics, Austrian Academy of Sciences,
Nikolsdorfergasse 18, A-1050 Vienna, Austria,\\Faculty of Physics,
University of Vienna, Boltzmanngasse 5, A-1090 Vienna, Austria,
and\\D.~V.~Skobeltsyn Institute of Nuclear Physics, Moscow State
University, 119991, Moscow, Russia\\E-mail:
\email{dmitri\_melikhov@gmx.de}} 

\abstract{The local-duality formulation of QCD sum rules allows
for the prediction of hadronic form factors without knowledge of
the subtle details of their structure. With the aid of this
formalism, we take a fresh look at the behaviours of the
charged-pion elastic form factor and of the form factors entering
in the transitions of the ground-state neutral unflavoured
pseudoscalar mesons $\pi^0,\eta,\eta'$ to~one~real and one virtual
photon within a broad range of momentum transfers $Q^2.$ The
uncertainties induced by the approximations inherent to this
local-duality approach are estimated by studying, in parallel to
QCD, quantum-mechanical potential models, where the exact form
factors, obtained by solving the Schr\"odinger equation, may be
compared with the corresponding local-duality sum-rule results.
For $Q^2\ge5$--$6\;\mbox{GeV}^2,$ we judge the predictions of the
simplest local-duality model to be reliable and expect their
accuracy to improve very fast with increasing $Q^2.$ The
large-$Q^2$ prediction for the~pion elastic form factor should be
approached already at moderate momentum transfer
$Q^2\approx4$--$8\;\mbox{GeV}^2;$ large deviations from its
local-duality behaviour for $Q^2=20$--$50\;\mbox{GeV}^2,$
predicted by some hadron-structure models, seem rather unlikely.
The $(\eta,\eta')\to\gamma\,\gamma^*$ form factors deduced from
the simplest local-duality approach exhibit excellent agreement
with experiment. In startling contrast, {\sc BaBar} measurements
of the $\pi^0\to\gamma\,\gamma^*$ form factor imply local-duality
violations which even rise with~$Q^2.$}

\FullConference{The XXth International Workshop High Energy
Physics and Quantum Field Theory\\September 24 -- October 1,
2011\\Sochi, Russia}

\begin{document}\section{Introduction: Motivation and Incentive
for Reconsidering a Long-Standing Issue} {\em QCD sum rules\/} aim
to predict the characteristic features of ground-state hadrons
(their masses, decay constants, form factors, etc.) from the
underlying quantum field theory of strong interactions, quantum
chromodynamics (QCD), by evaluating matrix elements of suitably
chosen operators both on the level of hadrons and on the level of
the QCD degrees of freedom quarks and gluons. Wilson's {\em
operator product expansion\/} allows for the conversion of these
{\em nonlocal\/} operators into series of local operators. By this
process the QCD-level matrix elements receive both perturbative
contributions as well as non-perturbative contributions involving
universal quantities called vacuum \mbox{condensates. In} order to
{\em suppress\/} the contributions of hadronic excitations and
continuum and to {\em remove\/} subtraction terms, {\em Borel
transformations\/} to new variables, dubbed as the Borel mass
parameters, are performed. Representing the perturbative
contributions to our QCD-level matrix elements in form of
dispersion integrals over corresponding spectral densities allows
us to bypass our ignorance about higher states by invoking the
concept of {\em quark--hadron duality\/}: beyond some {\em
effective thresholds\/} the perturbative QCD contributions and the
expressions of hadron excitations and continuum are assumed to
cancel. The outcome of these steps are sum rules relating QCD
parameters to observable hadron properties. In the limit of {\em
infinitely large\/} Borel mass parameters, all non-perturbative
QCD contributions vanish and we are left with what is known as
local-duality (LD) form of QCD sum rules, rendering possible to
derive features of ground-state hadrons from perturbative QCD and
our effective-threshold ideas.

Recently, we applied the LD sum-rule formalism to reanalyze both
the elastic form factor of the pion \cite{blm2011} and the form
factor that describes the transition $P\to\gamma\,\gamma^*$ of
some light neutral~pseudoscalar meson $P=\pi^0,\eta,\eta'$ to a
real photon $\gamma$ and a virtual photon $\gamma^*$
\cite{lm2011}. One particularly attractive~feature of the LD
sum-rule approach is the possibility to extract predictions for
hadron form factors without knowledge of all subtle details of the
structure of the hadronic bound states and to consider different
hadrons on an equal footing. Here, we take a retrospective look
from bird's eye view at our findings: After recalling, for the
example of the pion, the rather well-known basic features of the
LD sum-rule approach to pseudoscalar-meson form factors, in order
to get an idea (or even rough estimate) of the accuracy to be
expected for real-life mesons described by QCD sum rules we make a
brief and in the meanwhile well-established sidestep to their
quantum-mechanical analogues as a means to examine the
uncertainties induced by modeling the impact of higher hadronic
states in a rather na\"ive fashion. Then, equipped with sufficient
confidence in the reliability of the adopted LD approximation for
the effective thresholds, we discuss, in turn, the $\pi$ elastic
and $\left(\pi^0,\eta,\eta'\right)\to\gamma\,\gamma^*$ transition
form~factors.

\section{Dispersive Three-Point QCD Sum Rules in the Limit of Local
Duality \cite{ld}}The basic objects exploited here for the
investigation of the behaviour of form factors $F(Q^2)$ as
functions of the involved momentum transfers squared,
$Q^2=-q^2\ge0,$ are {\em three-point functions\/}, the vacuum
correlator of one vector and two axialvector currents, with {\em
double spectral density\/}~$\Delta_{\rm pert},$~for the elastic
form factor $F_\pi(Q^2)$ and the vacuum correlator of one
axialvector and two vector currents, with {\em single spectral
density\/} $\sigma_{\rm pert},$ for the transition form factor
$F_{\pi\gamma}(Q^2),$ satisfying the LD sum
rules\begin{equation}\label{Fld}
F_\pi(Q^2)=\frac{1}{f_\pi^2}\int\limits_0^{s_{\rm eff}(Q^2)}
\hspace{-1.7ex}{\rm d}s_1\hspace{-1ex}\int\limits_0^{s_{\rm
eff}(Q^2)}\hspace{-1.7ex}{\rm d}s_2\,\Delta_{\rm
pert}(s_1,s_2,Q^2)\ ,\qquad
F_{\pi\gamma}(Q^2)=\frac{1}{f_\pi}\int\limits_0^{\bar s_{\rm
eff}(Q^2)}\hspace{-1.7ex}{\rm d}s\,\sigma_{\rm pert}(s,Q^2)\
.\end{equation}Here, $f_\pi$ is the charged-pion decay constant:
$f_\pi=130\;\mbox{MeV}.$ Now all details of the non-perturbative
dynamics are encoded in the effective thresholds $s_{\rm
eff}(Q^2)$ and $\bar s_{\rm eff}(Q^2)$ that enter as upper
endpoints.

We take the liberty of introducing the notion of an {\em
equivalent effective threshold\/}, defined by the requirement that
the use of this quantity as effective threshold in the appropriate
dispersive sum rule --- such as the LD representatives of
Eq.~(\ref{Fld}) --- reproduces for the form factor under
consideration either given experimental data or a particular
theoretical prediction exactly. With such powerful tool at our
disposal, we are able to quantify our observations and make our
conclusions much~more~clear.

Within perturbation theory, the spectral densities $\Delta_{\rm
pert}(s_1,s_2,Q^2)$ and $\sigma_{\rm pert}(s,Q^2)$ are derived as
series expansions in powers of the strong coupling $\alpha_{\rm
s}$ by evaluating the relevant Feynman diagrams:
\begin{eqnarray}\Delta_{\rm pert}(s_1,s_2,Q^2)&=&\Delta^{(0)}_{\rm
pert}(s_1,s_2,Q^2)+\alpha_{\rm s}(Q^2)\,\Delta^{(1)}_{\rm
pert}(s_1,s_2,Q^2)+O(\alpha_{\rm s}^2)\ ,\nonumber\\\sigma_{\rm
pert}(s,Q^2)&=&\sigma^{(0)}_{\rm pert}(s,Q^2)+\alpha_{\rm
s}(Q^2)\,\sigma^{(1)}_{\rm pert}(s,Q^2)+O(\alpha_{\rm s}^2)\
.\label{Eq:SF}\end{eqnarray}As far as their aspects relevant for
our present purposes are concerned, the theoretical status of
these spectral densities may be summarized as follows. In the
double spectral density $\Delta_{\rm pert}(s_1,s_2,Q^2),$ for
fixed $s_{1,2}$ and large momentum transfers $Q^2,$ the one-loop
contribution $\Delta^{(0)}_{\rm pert}(s_1,s_2,Q^2)$ vanishes like
$\Delta^{(0)}_{\rm pert}(s_1,s_2,Q^2)\propto1/Q^4$ and the
two-loop contribution $\Delta^{(1)}_{\rm pert}(s_1,s_2,Q^2)$
approaches the behaviour \cite{bo}$$\Delta^{(1)}_{\rm
pert}(s_1,s_2,Q^2)\xrightarrow[Q^2\to\infty]{}\frac{1}{2\pi^3\,Q^2}\
;$$in other words, in the limit $Q^2\to\infty$ the lowest-order
term decays faster than the next-to-lowest term. In the single
spectral density $\sigma_{\rm pert}(s,Q^2),$ the two-loop
correction $\sigma^{(1)}_{\rm pert}(s,Q^2)$ has been proven
\cite{2loop} to vanish identically: $\sigma^{(1)}_{\rm
pert}(s,Q^2)\equiv0.$ Higher-order radiative corrections have not
yet been calculated.

With the required spectral densities available at least up to some
order of perturbation theory,~as soon as the dependencies of the
effective thresholds $s_{\rm eff}(Q^2)$ and $\bar s_{\rm
eff}(Q^2)$ on the momentum transfer $Q^2$ have been found, the
form factors of interest can be easily extracted from the LD sum
rules~(\ref{Fld}). Factorization theorems for hard form factors
\cite{brodsky}, allowing for separation of the dynamics into
short- and long-distance contributions, establish the asymptotic
behaviour of the form factors for large $Q^2$:
$$Q^2\,F_\pi(Q^2)\xrightarrow[Q^2\to\infty]{}8\pi\,\alpha_{\rm
s}(Q^2)\,f_\pi^2\ ,\qquad
Q^2\,F_{\pi\gamma}(Q^2)\xrightarrow[Q^2\to\infty]{}\sqrt{2}\,f_\pi\
.$$The sum rules (\ref{Fld}) with the spectral functions
(\ref{Eq:SF}) reproduce, at $O(\alpha_{\rm s}^2)$ accuracy, this
behaviour~if\begin{equation}\lim_{Q^2\to\infty}\hspace{-.7ex}s_{\rm
eff}(Q^2)=\lim_{Q^2\to\infty}\hspace{-.7ex}\bar s_{\rm
eff}(Q^2)={4\pi^2\,f_\pi^2}\approx0.671\;\mbox{GeV$^2$}\label{Eq:AET}
\end{equation}holds. The remaining task is to determine the behaviour
of the effective thresholds at finite values of $Q^2.$
Unfortunately, as analyzed in detail in Refs.~\cite{lms1}, the
formulation of a reliable criterion~for~fixing a threshold poses a
somewhat delicate problem as, for finite $Q^2,$ the effective
thresholds $s_{\rm eff}(Q^2)$~and $\bar s_{\rm eff}(Q^2)$ cannot
be assumed to be equal to their asymptotes (\ref{Eq:AET}); rather,
they will depend on $Q^2$ and, generally, differ from each other
\cite{lms2}. A very simple idea is to assume that the use of their
asymptotic values provides a meaningful approximation also at
moderate but not too small momentum transfer: $s_{\rm
eff}(Q^2)=\bar s_{\rm eff}(Q^2)={4\pi^2\,f_\pi^2}.$ This choice
defines a straightforward albeit rather na\"ive LD model
\cite{ld}. It goes without saying that such crude approximations
to the effective thresholds may be well~suited to reproduce the
overall trend but can hardly account for any subtle detail of
confinement dynamics.

\section{Exact and Local-Duality Form Factors in
Quantum-Mechanical Potential Models}The (quantum-field-theoretic)
LD sum-rule approach to bound-state form factors may be easily
carried over to quantum mechanics. Within the latter framework,
the features of any bound state can be obtained with, in
principle, arbitrarily high precision from the related solution of
the Schr\"odinger equation for the Hamiltonian governing the
dynamics of the system under consideration. Therefore,
quantum-mechanical potential models constitute an ideal test
ground for estimating the significance of LD models that employ
for the effective thresholds entering in the adopted sum rules the
constant limits fixed by some asymptotic behaviour at
experimentally accessible lower momentum transfers. For this very
reason, we examine quantum-mechanical potential models defined by
Hamiltonians $H$ which must incorporate, for the study of the
elastic form factor, confining {\em and\/} Coulomb interactions
($\eta=1$) but, for the investigation of the transition form
factor, merely confining interactions ($\eta=0$):
$$H=\frac{\mathbf{k}^2}{2\,m}+V_{\rm
conf}(r)-\eta\,\frac{\alpha}{r}\ ,\qquad V_{\rm
conf}(r)=\sigma_n\,(m\,r)^n\ ,\qquad r\equiv|\mathbf{x}|\ ,\qquad
n=2,1,1/2\ .$$ We ensure a realistic description of mesons by
adopting for our numerical analysis parameter~values appropriate
for hadron physics: $m=0.175\;\mbox{GeV}$ for the reduced mass of
light constituent quarks~and $\alpha=0.3$ for the coupling
strength $\alpha$ of the Coulomb interaction term. For the
confining interactions, we consider several power-law potential
shapes $V_{\rm conf}(r),$ adjusting the associated coupling
strengths $\sigma_n$ such that in each case the Schr\"odinger
equation predicts the same value
$\psi(0)=0.078\;\mbox{GeV}^{3/2}$~for the ground-state wave
function $\psi$ at the origin: $\sigma_2=0.71\;\mbox{GeV},$
$\sigma_1=0.96\;\mbox{GeV}$ and $\sigma_{1/2}=1.4\;\mbox{GeV}.$
Then, the size of the lowest-lying bound state is about 1\;fm and
thus of typical hadronic dimensions.

\begin{figure}[h]\begin{center}\begin{tabular}{cc}
\includegraphics[width=7.147cm]{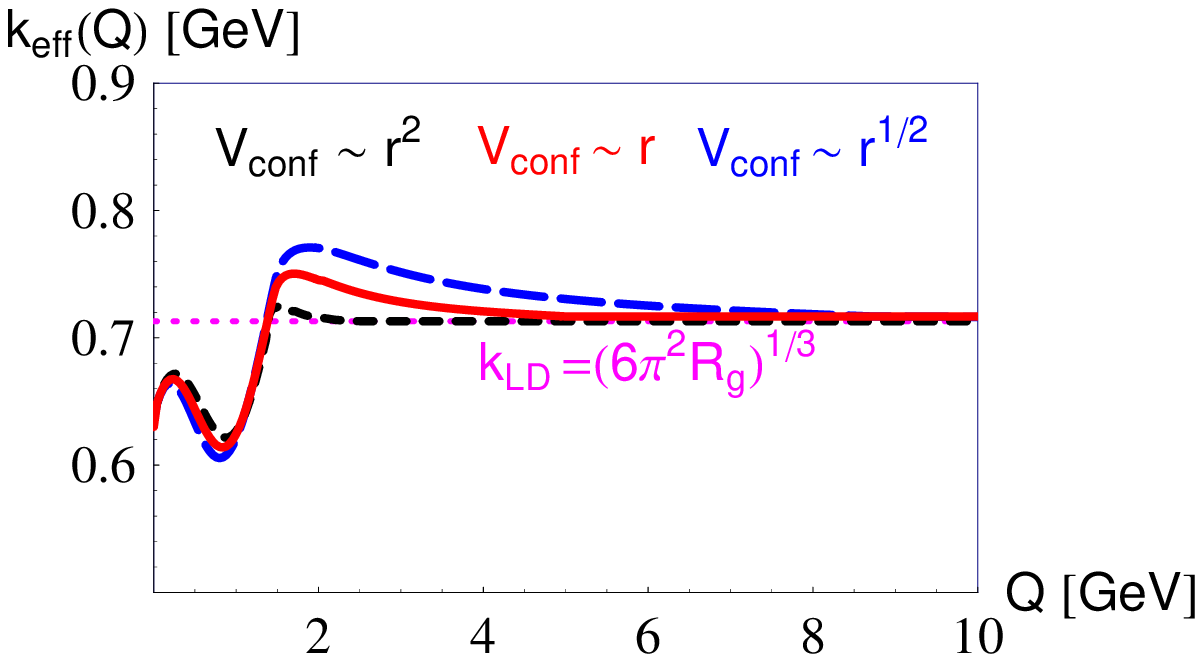}&
\includegraphics[width=7.147cm]{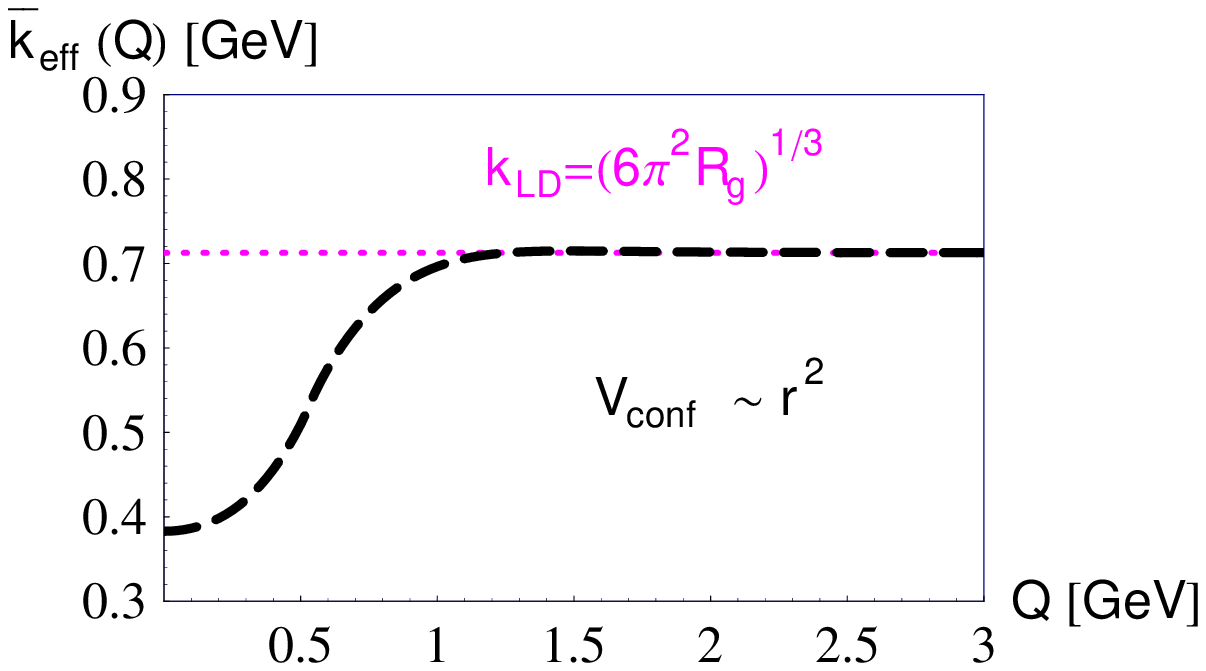}\end{tabular}
\caption{\label{Plot:1}Exact quantum-mechanical effective
thresholds for elastic (left) and transition (right) form
factors.}\end{center}\end{figure}

With the numerically exact solution of the Schr\"odinger equation
at hand, we are in a position to confront the form factors arising
thereof with corresponding predictions of the
\mbox{quantum-mechanical} counterparts of the LD QCD sum rules
(\ref{Fld}), which involve effective thresholds $k_{\rm eff}(Q)$
and $\bar k_{\rm eff}(Q),$ respectively. As in the QCD case, the
asymptotic behaviour of the elastic and transition form factors in
the limit of infinitely large momentum transfer $Q$ may be derived
from factorization theorems \cite{brodsky}. In terms of the
ground-state decay constant $R_{\rm g}\equiv|\psi(0)|^2,$ this
asymptotic behaviour~is~guaranteed if the effective thresholds
fulfill $k_{\rm eff}(Q\to\infty)=\bar k_{\rm
eff}(Q\to\infty)=(6\pi^2\,R_{\rm g})^{1/3}.$ Figure \ref{Plot:1}
shows~that~the LD model $k_{\rm eff}(Q)=\bar k_{\rm
eff}(Q)=(6\pi^2\,R_{\rm g})^{1/3}$ approximates independently of
the confining potential in use the exact effective thresholds
yielding the true form factors with improving accuracy, starting
for the elastic form factor at $Q^2\approx5$--$8\;\mbox{GeV}^2$
and for the transition form factor at some~even~lower~$Q^2$~value.

\section{The Pion Elastic Form Factor \cite{blm2011}}
\begin{figure}[hbt]\begin{center}\begin{tabular}{c}
\includegraphics[width=7.5cm]{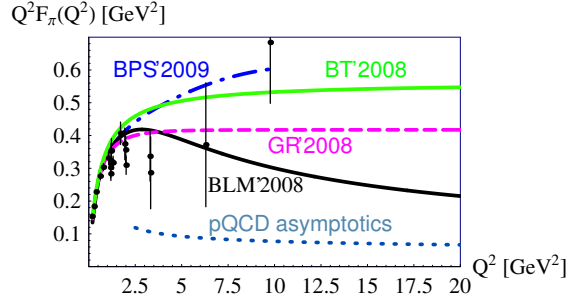}\end{tabular}
\caption{\label{Plot:2}Pion elastic form factor $F_\pi(Q^2)$:
experimental data \cite{data_piff} and some recent theoretical
findings \cite{blm2011,recent}.}\end{center}\end{figure}

The pion
belongs, beyond doubt, to the best-studied mesons. Nevertheless,
the one or the other of its most important properties still cannot
seriously be claimed to be sufficiently well
understood.\footnote{Of course, whenever some problem in the
treatment of any of the ground-state pseudoscalar mesons is
encountered, as a kind of automatic reflex-like response one may
be tempted to blame within QCD the pseudo-Goldstone-boson nature
of the particle for preventing us from acquiring a satisfactory
understanding. Nevertheless, all comprehensive approaches should
be expected to be able to deal with this sort of ``inconvenience''
and to ultimately incorporate such crucial features.} Figure
\ref{Plot:2} displays a snapshot of the present status of the
pion's electromagnetic or elastic form factor $F_\pi(Q^2)$ from
both the experimental \cite{data_piff} and the theoretical
\cite{blm2011,recent} points of view. Obviously, there is ample
room for controversy, but no consensus on $F_\pi(Q^2)$ for
momentum~transfers $Q^2\approx5$--$50\;\mbox{GeV}^2.$

\begin{figure}[hbt]\begin{center}\begin{tabular}{cc}
\includegraphics[width=7.147cm]{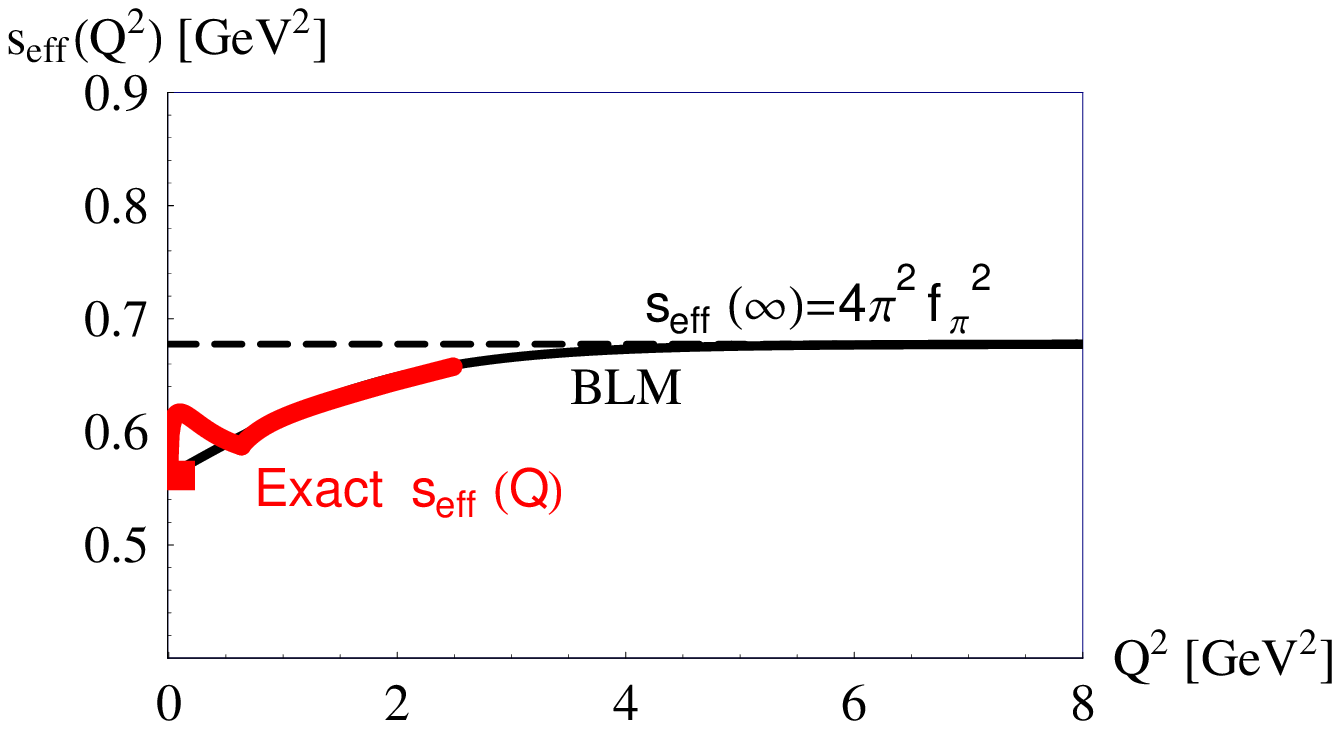}&
\includegraphics[width=7.147cm]{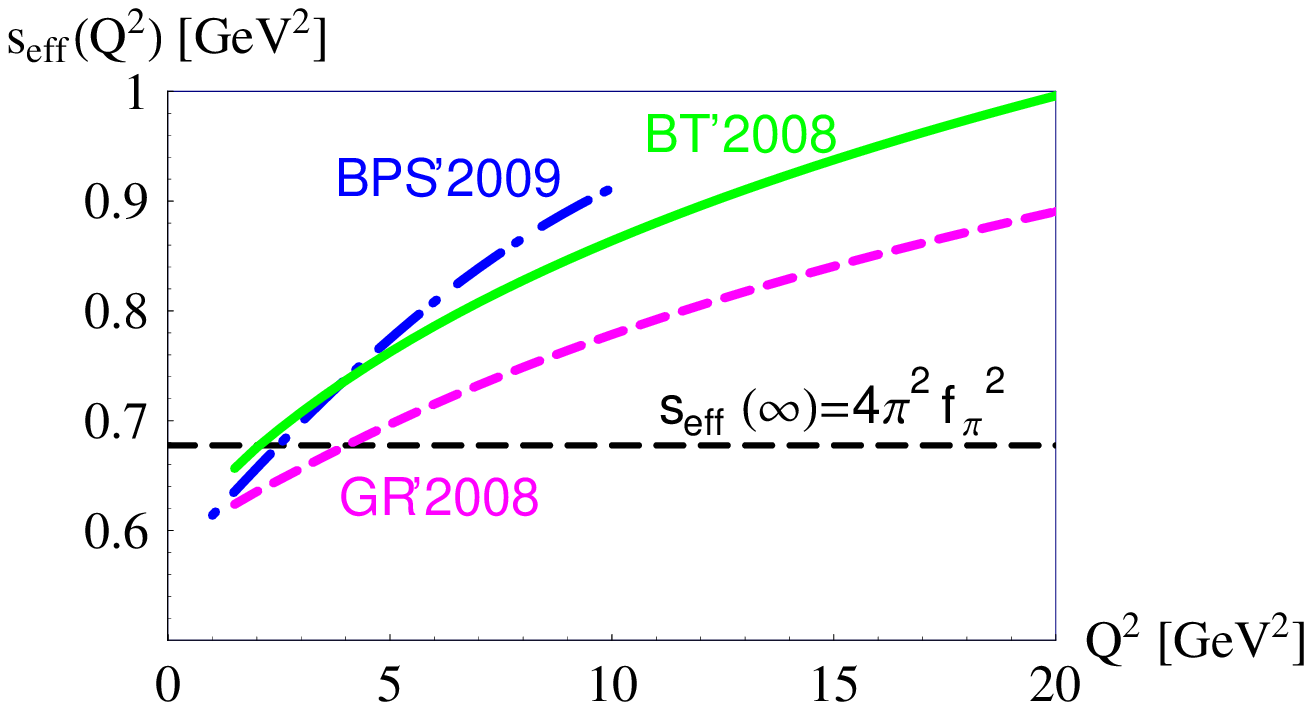}
\end{tabular}\caption{\label{Plot:3}Parametrization of
the effective threshold $s_{\rm eff}(Q^2)$ by an improved LD model
\cite{blm2011} (labelled BLM) vs.\ exact behaviour (red) of the
equivalent effective threshold extracted from experimental data
\cite{data_piff} (left),~and equivalent effective thresholds
corresponding to the theoretical results for $F_\pi$
\cite{blm2011,recent} depicted in Fig.~\protect\ref{Plot:2}
(right).}\end{center}\end{figure}

In order to cast some light onto these disquieting puzzles, Figure
\ref{Plot:3} depicts our translation~of the findings summarized in
Fig.~\ref{Plot:2} to equivalent effective thresholds $s_{\rm
eff}(Q^2)$ calculated back from either experimental data or
theoretical predictions for $F_\pi(Q^2)$: the exact effective
threshold extracted~from the data is compatible with the
assumption that the LD limit is approached at rather low $Q^2$
whereas, contrary to quantum physics, theory seems not to care
about local duality, at least for $Q^2\le20\;\mbox{GeV}^2.$

Rather precise measurements may be expected from JLab after the
12\;GeV upgrade~of~CEBAF.

\section{The $(\pi^0,\eta,\eta')\to\gamma\,\gamma^*$ Transition
Form Factors \cite{lm2011}}In order to consolidate our concerns
and to substantiate our confusions, we discuss the $\eta$~and
$\eta'$ transitions $(\eta,\eta')\to\gamma\,\gamma^*$ before
turning to the controversial issue of the pion's transition
$\pi^0\to\gamma\,\gamma^*.$

\subsection{Form Factors for the Transitions
$(\eta,\eta')\to\gamma\,\gamma^*$}The two isoscalar mesons $\eta$
and $\eta',$ having the same $J^{PC}$ quantum numbers, are
mixtures of {\em all\/} light quarks. In the flavour basis, the
mixing of the non-strange and strange contributions~is~given~by
$$|\eta\rangle=\left|\frac{\bar uu+\bar
dd}{\sqrt2}\right\rangle\cos\phi-|\bar ss\rangle\sin\phi\
,\qquad|\eta'\rangle=\left|\frac{\bar uu+\bar
dd}{\sqrt2}\right\rangle\sin\phi+|\bar ss\rangle\cos\phi\ ,$$with
mixing angle $\phi\approx39.3^\circ;$ see, {\em e.g.\/},
Refs.~\cite{anisovich,feldmann}. The form factors reflect this
flavour structure:
$$F_{\eta\gamma}(Q^2)=\frac{5\,F_{n\gamma}(Q^2)}{3\,\sqrt2}\cos\phi
-\frac{F_{s\gamma}(Q^2)}{3}\sin\phi\ ,\qquad
F_{\eta'\gamma}(Q^2)=\frac{5\,F_{n\gamma}(Q^2)}{3\,\sqrt2}\sin\phi
+\frac{F_{s\gamma}(Q^2)}{3}\cos\phi\ .$$Here, the non-strange and
$(\bar ss)$ components $F_{n\gamma}(Q^2)$ and $F_{s\gamma}(Q^2)$
of the LD form factors are given~by
$$F_{n\gamma}(Q^2)=\frac{1}{f_n}\int\limits_0^{\bar s_{\rm
eff}^{(n)}(Q^2)}\hspace{-1.7ex}{\rm d}s\,\sigma^{(n)}_{\rm
pert}(s,Q^2)\ ,\qquad
F_{s\gamma}(Q^2)=\frac{1}{f_s}\int\limits_{4\,m_s^2}^{\bar s_{\rm
eff}^{(s)}(Q^2)}\hspace{-1.7ex}{\rm d}s\,\sigma^{(s)}_{\rm
pert}(s,Q^2)\ ,$$where $\sigma^{(n)}_{\rm pert}$ and
$\sigma^{(s)}_{\rm pert}$ label the single spectral density
$\sigma_{\rm pert}$ of Eq.~(\ref{Fld}) with the
corresponding~quark, $n=u,d$ or $s,$ propagating in the loop; each
component utilizes an effective threshold of its own
\cite{feldmann}:$$\bar s_{\rm eff}^{(n)}=4\pi^2\,f_n^2\ ,\qquad
f_n\approx1.07\,f_\pi\ ,\qquad\bar s_{\rm
eff}^{(s)}=4\pi^2\,f_s^2,\qquad f_s\approx1.36\,f_\pi\ .$$In our
numerical calculations, we adopt $m_u=m_d=0$ and
$m_s=100\;\mbox{MeV}$ for the light-quark~masses.

\begin{figure}[hbt]\begin{center}\begin{tabular}{cc}
\includegraphics[width=7.147cm]{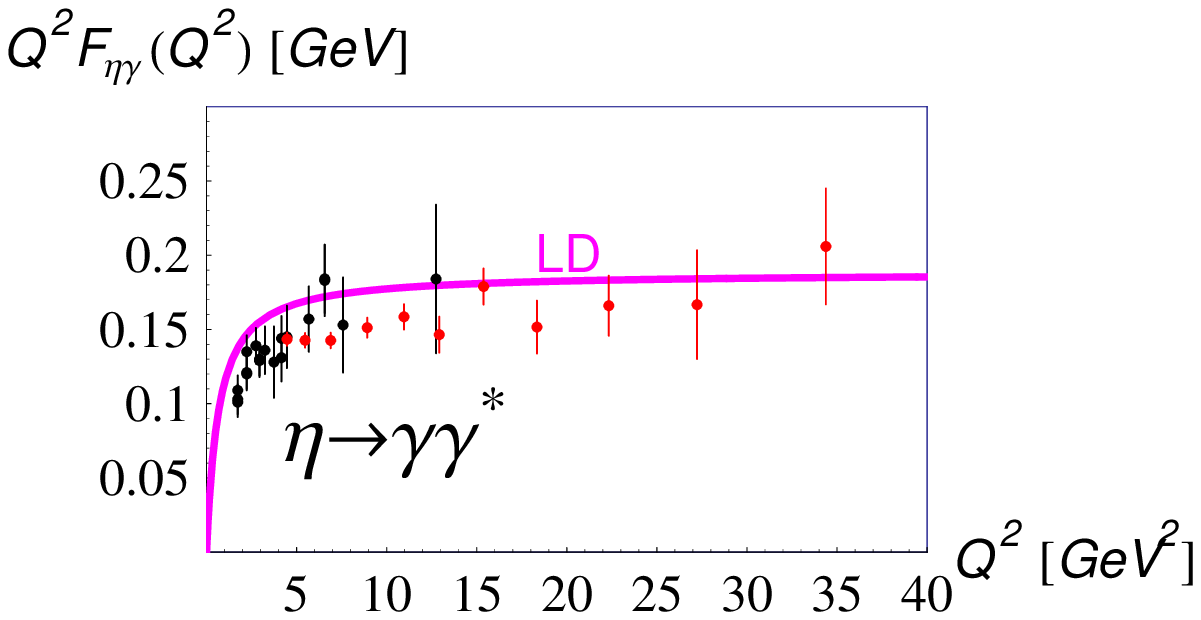}&
\includegraphics[width=7.147cm]{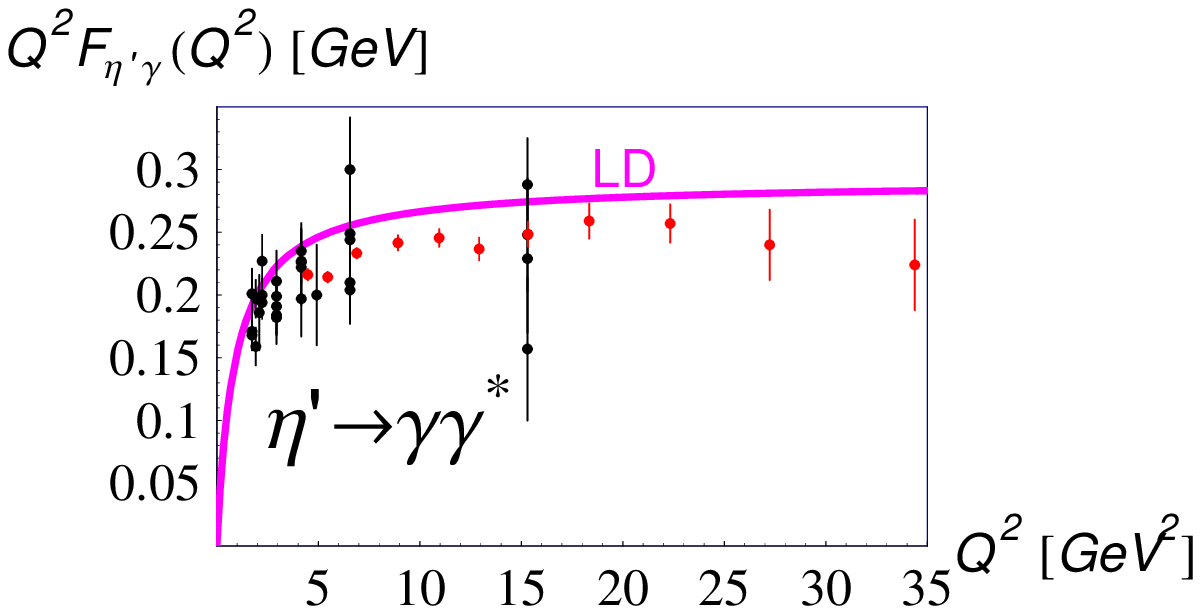}
\end{tabular}\caption{\label{Plot:4}Transition form factors
$F_{(\eta,\eta')\gamma}(Q^2)$: for $\eta$ and $\eta'$ the LD model
fits the experimental data \cite{cello-cleo,babar1}.}\end{center}
\end{figure}

According to all our experience gained by in-depth investigations
of the LD sum-rule approach within quantum mechanics, this
straightforward but admittedly not too sophisticated LD~framework
may not perform really well for low momentum transfers $Q^2,$
where, as a brief look at Fig.~\ref{Plot:3} reveals, the exact
effective threshold is below the constant LD effective threshold
inferred from~the~large-$Q^2$ form-factor behaviour. However, for
larger momentum transfer the simple~quantum-mechanical~LD model
entails accurate predictions for form factors. Figure \ref{Plot:4}
shows that, for both $\eta$ and $\eta'$ transition form factors,
we find the anticipated overall agreement between the LD
predictions and~experiment.

\subsection{Form Factor for the Transition
$\pi^0\to\gamma\,\gamma^*$}
\begin{figure}[hbt]\begin{center}\begin{tabular}{cc}
\includegraphics[width=7.147cm]{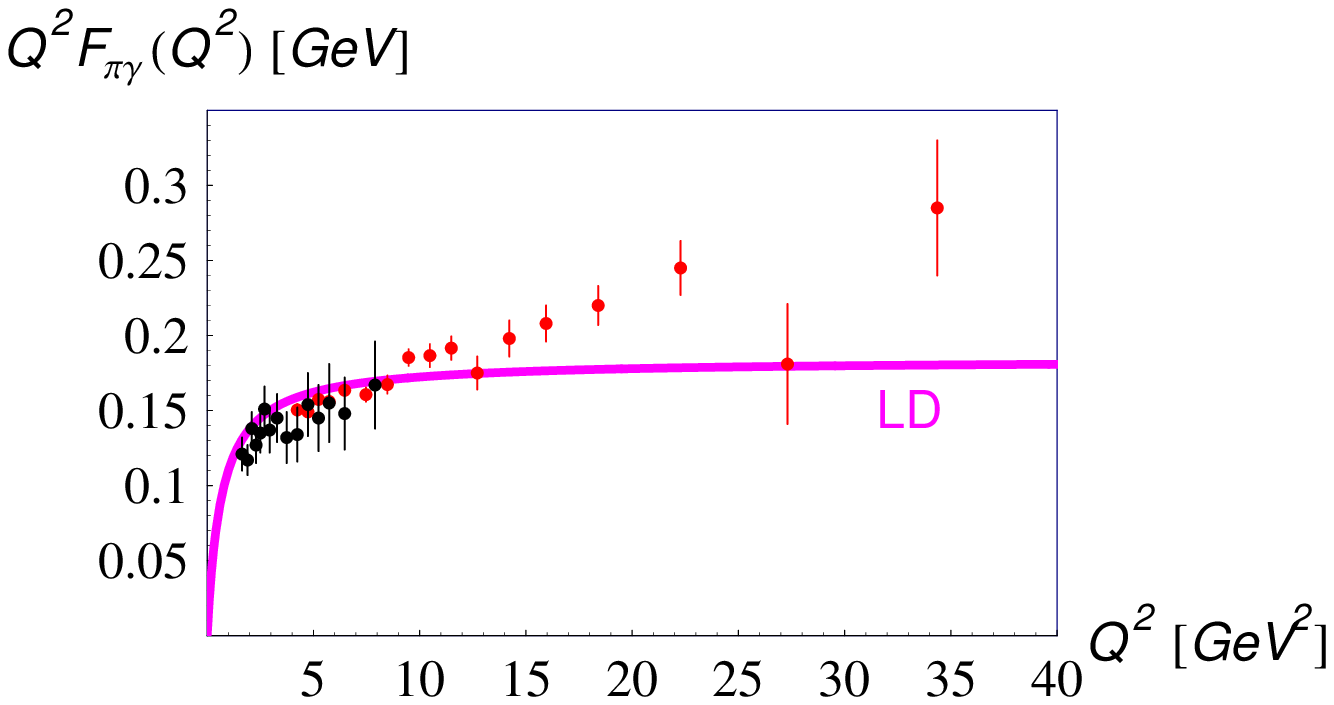}&
\includegraphics[width=7.147cm]{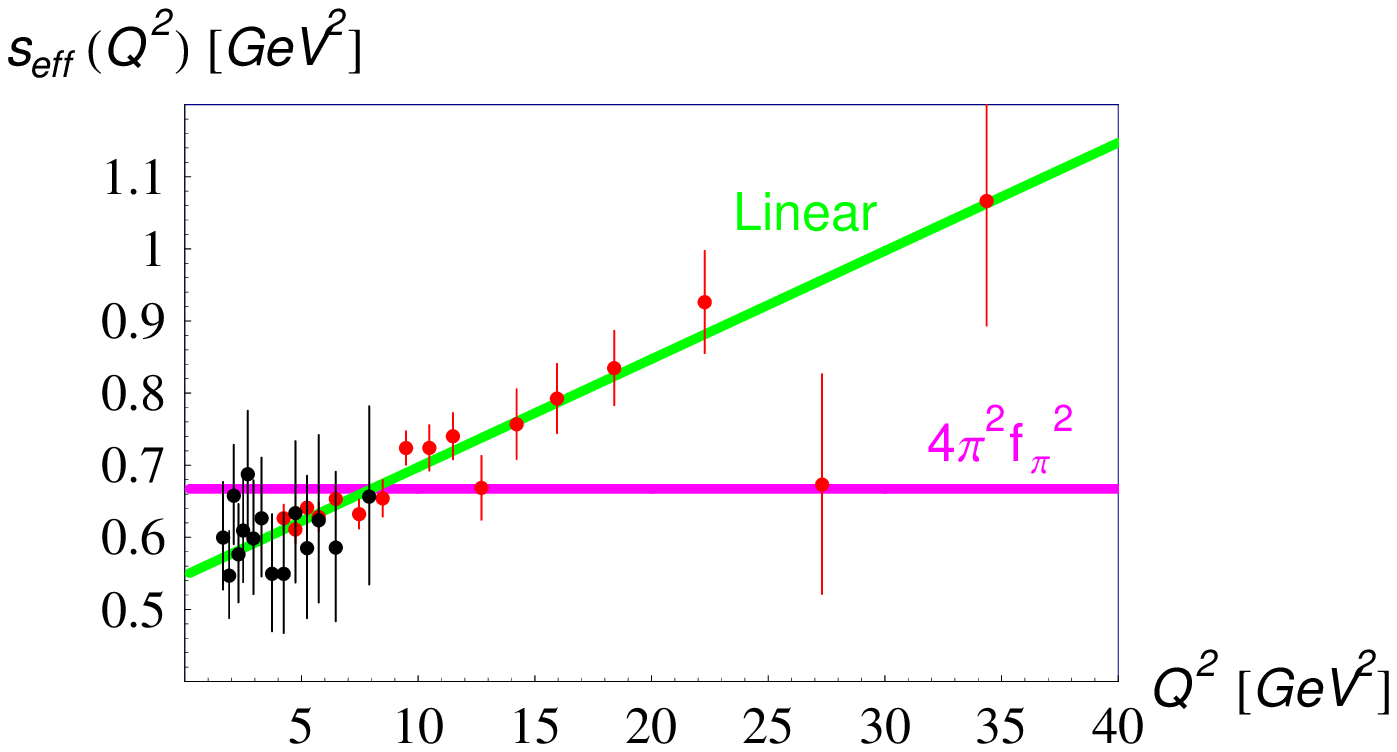}
\end{tabular}\caption{\label{Plot:5}Form factor $F_{\pi\gamma}(Q^2)$
for the pion transition $\pi^0\to\gamma\,\gamma^*$: some
experimental data \cite{cello-cleo,babar}, at least the {\sc
BaBar} data (red dots), apparently diverge from our LD prediction;
this unexpected behaviour is reflected by the equivalent effective
threshold $\bar s_{\rm eff}(Q^2)$ exhibiting a linear rise with
$Q^2$ instead of approaching~its~LD~limit.}\end{center}
\end{figure}

In view of the undeniable successes of
the LD model in the case of the $\pi$ elastic form factor and of
the $\eta$ and $\eta'$ transition form factors, its failure in the
case of the $\pi^0$ transition form factor $F_{\pi\gamma}(Q^2)$ is
all the more surprising. Figure \ref{Plot:5} displays how markedly
the LD prediction for $F_{\pi\gamma}(Q^2)$ misses the {\sc BaBar}
data \cite{babar}. This becomes even more manifest by the linear
rise with $Q^2$ of the corresponding equivalent effective
threshold $\bar s_{\rm eff}(Q^2),$ which, at least in the region
up to $Q^2\approx40\;\mbox{GeV}^2,$ exhibits no tendency of
approaching its LD limit (\ref{Eq:AET}). This intriguing puzzle
still awaits a compelling~solution.

\section{Summary: Findings and Conclusions}By reconsidering the
dependence of the pion elastic \cite{blm2011} and
$\pi^0,\eta,\eta'$ transition \cite{lm2011} form factors on the
momentum transfer $Q^2$ using QCD sum rules in LD limit, we gain
highly interesting insights:\begin{description}\item[Pion elastic
form factor:] Transferring the outcomes of our quantum-mechanical
analysis to QCD, we expect the simple LD model to be applicable
with increasing accuracy for $Q^2\ge4$--$8\;\mbox{GeV}^2$
irrespective of the adopted confining interactions. For realistic
confining interactions, this LD model reproduces the elastic form
factor for $Q^2\ge20$--$30\;\mbox{GeV}^2$ with high precision.
Accurate measurements of this form factor at small $Q^2$ suggest
that assuming for the effective threshold its LD limit already at
rather low $Q^2=5$--$6\;\mbox{GeV}^2$ may constitute a reasonable
approximation. Hence, large deviations from this LD limit at
$Q^2=20$--$50\;\mbox{GeV}^2$ must be regarded as
unlikely.\item[Transition form factors for $\pi^0,\eta,\eta'$:]
Our observations in quantum mechanics can be understood as hints
that, for bound states of typical hadron extensions, the LD
approach should work well for $Q^2$ larger than a few
$\mbox{GeV}^2,$ and it indeed does for the
$\eta\to\gamma\,\gamma^*$ and $\eta'\to\gamma\,\gamma^*$
form~factors. However, a recent measurement of the form factor for
the neutral-pion transition $\pi^0\to\gamma\,\gamma^*$ by the {\sc
BaBar} experiment \cite{babar} implies a violation of local
duality which even grows with $Q^2,$ at least up to $Q^2$ as high
as $40\;\mbox{GeV}^2.$ Within the LD sum-rule formalism
(\ref{Fld}), such behaviour of a transition form factor cannot be
accommodated by a {\em constant\/} equivalent effective threshold
but must be described by a linear $Q^2$-dependence of $\bar s_{\rm
eff}(Q^2);$ a convincing explanation of this has yet to be found.
This conclusion enjoys full agreement with the findings of
Refs.~\cite{roberts,mikhailov}.\end{description}

\acknowledgments{D.M.\ is grateful for financial support by the
Austrian Science Fund (FWF), project no.~P22843.}

\end{document}